\documentclass[5p]{elsarticle}

\usepackage{lineno,hyperref}
\usepackage{xcolor}
\usepackage{hyperref}

\journal{Acta Astronautica}

\biboptions{sort&compress}
\begin{document}

\begin{frontmatter}

\title{Evaluating potential landing sites for the Artemis III mission using a multi-criteria decision making approach}

\author[A1]{Eloy Peña-Asensio}
\author[A2]{Álvaro-Steve Neira-Acosta}
\author[A3]{Juan Miguel Sánchez-Lozano\corref{mycorrespondingauthor}}
\cortext[mycorrespondingauthor]{Corresponding author}
\ead{juanm.sanchez@upct.es}




\address[A1]{Dipartimento di Scienze e Tecnologie Aerospaziali, Politecnico di Milano (PoliMi), 20156, Milano, Italy}
\address[A2]{Valencian International University (VIU), 46002, València, Spain}
\address[A3]{Department of Structures, Construction, and Graphic Expression, Universidad Politécnica de Cartagena (UPCT), 30202, Cartagena, Spain}


\begin{abstract}

The selection of a landing site within the Artemis Exploration Zone (AEZ) involves multiple factors and presents a complex problem. This study evaluates potential landing sites for the Artemis III mission using a combination of Geographic Information Systems (GIS) and Multi-Criteria Decision Making (MCDM) methodologies, specifically the TOPSIS algorithm. By integrating topographic, illumination, and mineralogy data of the Moon, we assess 1247 locations that meet the Human Landing System (HLS) requirements within 13 candidate regions and Site 004 near the lunar south pole. Criteria considered include surface visibility, HLS-astronaut line of sight, Permanently Shadowed Regions (PSRs), sunlight exposure, direct communication with Earth, geological units, and mafic mineral abundance. Site DM2 (Nobile Rim 2), particularly the point at latitude 84°12'5.61" S (-84.20156°) and longitude 60°41'59.61" E (60.69989°), is the optimal location for landing. Sensitivity analysis confirms the robustness of our approach, validating the suitability of the best location despite the MCDM method employed and variations in criteria weightings to prioritize illumination and PSRs. This research demonstrates the applicability of GIS-MCDM techniques for lunar exploration and the potential benefits they can bring to the Artemis program.

\end{abstract}

\begin{keyword}
Lunar Exploration \sep Landing Site \sep Artemis III \sep Multi-Criteria Decision Making
\end{keyword}

\end{frontmatter}


\section{Introduction}

Neil Armstrong first walked on the Moon in July 1969 during the Apollo 11 mission \cite{NASA1969}. Over the past 55 years, interest in lunar exploration has grown, particularly with NASA's Artemis program. Unlike earlier missions that prioritized communication with Earth by landing in the equatorial zone of the lunar nearside \cite{NASA1970, NASA1971, NASA1972a, NASA1972b, NASA1973}, Artemis III intends to explore the lunar south pole.

The south pole of the Moon offers a unique opportunity to study Permanently Shadowed Regions (PSRs), which might contain water ice. These ice reserves will enable extended lunar missions, producing propellants for future explorations, and analyzing trapped volatiles \cite{NASA2020}. However, the Artemis Exploration Zone (AEZ), situated 6$^{\circ}$ latitude from the south pole, has a rugged and complex terrain that could impact astronaut Extra-Vehicular Activities (EVA) \cite{Kringetal2023, Johnson2024}. Hence, selecting an area that meets specific criteria is vital. These criteria include safe and level topography to reduce risks during landing and operations, direct communication with Earth for efficient data transmission and mission control, adequate sunlight for solar power generation, and maintaining operational temperatures for equipment, all while maximizing scientific returns \cite{NASAWeber2021LPI521261W}.

\begin{figure*}[h!]
\centering
\includegraphics[width=0.75\textwidth]{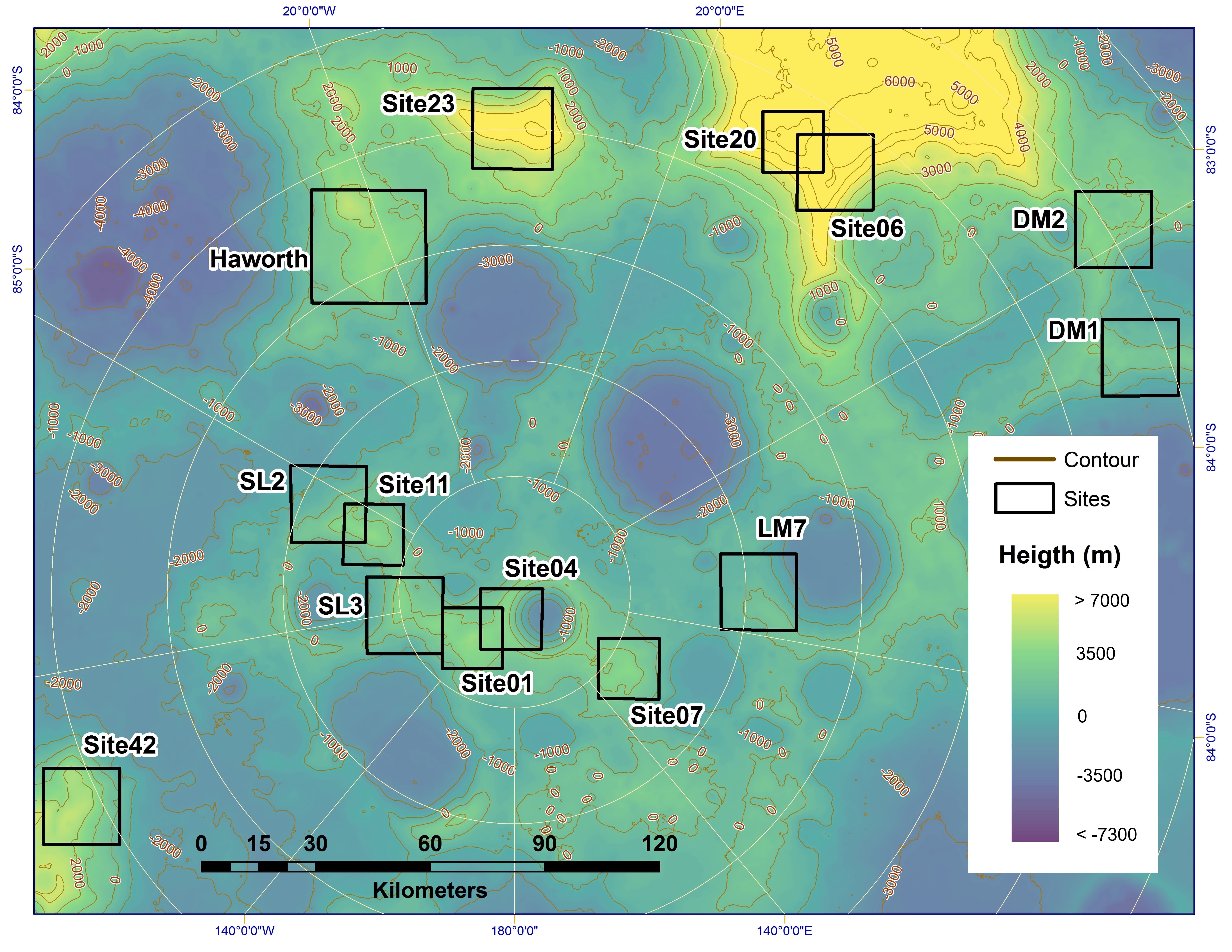}
\caption{Map of 13 candidate landing regions for Artemis III, in addition to Site 004. Each region measures approximately $\sim$15$\times$15 km. Terrain farther from the south pole, like Haworth crater, appears larger than it is due to the projection effect. Contours represent 1 km elevation intervals. Concentric rings at steps of 1$^{\circ}$ in latitude. }
\label{fig1}
\end{figure*}

Thirteen candidate regions of interest for Artemis III that meet the aforementioned criteria have been officially pre-selected\footnote{\url{https://www.nasa.gov/news-release/nasa-identifies-candidate\allowbreak-regions-for-landing-next-americans-on-moon/}}, see Figure~\ref{fig1}. Several studies have been published focusing on the geology and potential astronaut exploration within these sites \cite{2020Icar34813850B, Cannon2020ESS, 2020AdSpR661247G, Czaplinski2021PSJ, Lemelin2021PSJ, Bickel2022GeoRL, Kumari2022PSJ, Boazman2022PSJ, Bernhardt2022Icar, PENAASENSIO2024, Moriarty2024JGRE}. However, NASA has not yet selected a landing site for the Artemis III mission \cite{NASA2023}. This challenge involves selecting the optimal location among several alternatives, each influenced by multiple factors or criteria. Decision theory offers algorithms and techniques to address such problems. These are known as Multi-Criteria Decision Making (MCDM) methodologies \cite{Triantaphyllou2000}, which we apply in this work to rank the potential landing sites among these regions.

Since the introduction of MCDM methods in the latter half of the 20th century, various algorithms and techniques have been developed. Key examples include the DEcision-MAking Trial and Evaluation Laboratory (DEMATEL) \cite{Gabus1972}, Analytic Hierarchy Process (AHP) \cite{Saaty1980}, Technique for Elimination and Choice Translating Reality (ELECTRE) \cite{Roy1978}, Order of Preference by Similarity to Ideal Solution (TOPSIS) \cite{HwangandYoon1981}, Preference Ranking Organization Method for Enrichment Evaluation (PROMETHEE) \cite{Bransetal1984},  VIseKriterijumska Optimizacija I Kompromisno Resenje (VIKOR) \cite{Opri1998}. Recently developed approaches include Multi-Attributive Border Approximation area Comparison (MABAC) \cite{PAMUCAR2015}, Combined Compromise Solution (CoCoSo) \cite{Yazdani2019}, and Additive Ratio Assessment (ARAS) \cite{Zavadskas2010}. Each of these methodologies employs a decision matrix of alternatives and criteria, similar to the matrix needed for selecting the Artemis III mission landing site.

The current availability of detailed topographic models of the Moon \cite{Robinson2010SSRvLROC} facilitates the generation of databases of alternatives and criteria for landing site selection. These can be effectively managed using computer-based spatial visualization tools such as Geographic Information Systems (GIS). GIS software is an ideal complement to MCDM methodologies as it allows an easy generation of the decision matrix, which is the starting point of any multi-criteria analysis \cite{Jankowski1995}. In fact, the GIS-MCDM combination has been widely applied to solve various location problems on Earth \cite{CHANG2008, CHEN2010, SANCHEZLOZANO2013, SANCHEZLOZANO2014, TAHRI2015, ALGARNI2017, KHOSRAVI2019, SHAO2020, Rahimabadi2023, ROMERORAMOS2023, MOHAMEDetal2023, ALOUI2024, Zandi2024, Arabatzis2024}.

The TOPSIS method is often integrated with Geographic Information Systems (GIS) for several reasons. Its primary benefits include ease of computation, straightforward logic, and mathematical robustness \cite{wang2007}. Additionally, it can manage numerous alternatives based on criteria of varying nature and units of measurement. These attributes make TOPSIS particularly effective when various data layers from GIS software intersect at each pixel, identifying the alternatives to prioritize. The scientific literature indicates that the GIS-MCDM combination, although previously unused beyond terrestrial applications, is well-suited for addressing optimal location problems like the one in this study. Consequently, we utilize the GIS-TOPSIS combination to determine the optimal landing site for the Artemis III mission.

The structure of this study is as follows: Section \ref{methodology} describes the databases used and the methodology applied, Section \ref{alternatives} describes the generation of the potential landing locations, Section \ref{criteria} details the definition of the criteria employed for the evaluation, Section \ref{prioritization} shows the results, Section \ref{sensitivity} examines the robustness of the methodology through a sensitivity analysis, and finally Section \ref{conc} summarizes the conclusions drawn and potential future work.

\section{Methods and datasets}\label{methodology}

Our goal is to identify the optimal landing sites within specific regions of the AEZ using the TOPSIS method. This approach allows us to evaluate various point locations based on the following multiple criteria: visibility, PSRs, solar illumination, direct communication with Earth, geological units, and the abundance of mafic materials. The visibility maps are generated specifically for this study. The other maps utilized are based on previously produced and openly available products, as detailed later on.

The TOPSIS method, introduced by \cite{HwangandYoon1981}, is a popular MCDM technique used in engineering projects \cite{Mardani2015}. This approach calculates the relative closeness of each alternative to two hypothetical ideal solutions using Euclidean distance: the positive ideal solution and the negative ideal solution. According to TOPSIS, the optimal alternative is the one nearest to the positive ideal solution and furthest from the negative ideal solution. This relative closeness results in a ranking or prioritization of the alternatives. The principles of this MCDM technique are comprehensively detailed in the scientific literature \cite{Triantaphyllou2000}. In this work, we apply the TOPSIS method to prioritize the best landing site for the Artemis III mission.

The TOPSIS method involves the following steps:

\begin{itemize}
    \item Step 1. Construct a performance decision matrix composed of rows representing the alternatives \( A_i~(i=1,2,\ldots,m) \) to be evaluated and columns representing the criteria \( C_j~(j=1,2,\ldots,n) \) that influence the assessment process.
    
    \item Step 2. Normalize the decision matrix using the following expression:
\begin{equation}
\label{eqn:normalizing}
n_{ij} = \frac{x_{ij}}{\sqrt{\sum_{i=1}^{m}x_{ij}^2}},
\end{equation}
where \( n_{ij} \) represents elements in the normalized decision matrix \( N \).

    \item Step 3. Compute the weighted normalized decision matrix by multiplying the normalized decision matrix by the criteria weight vector:
\begin{equation}
\label{eqn:weighting}
v_{ij} = w_j \otimes n_{ij}, \quad j=1,\ldots,n, \quad i=1,\ldots,m,
\end{equation}
where \( w_j \) values are such that \( \sum_{j=1}^{n}w_j=1 \).

    \item Step 4. Identify the positive ideal \( A^+ \) and negative ideal \( A^- \) solutions, which depend on the nature of the criteria; for a benefit criterion, \( A^+ \) is the highest value, and for a cost criterion, it is the lowest, and the opposite applies for \( A^- \):
\begin{equation}
\label{eq:bigkey}
A^+ = \left\lbrace v^+_1,\ldots,v^+_n \right\rbrace = \left\lbrace
\begin{array}{ll}
\max_i \left\lbrace v_{ij}, j \in J \right\rbrace \\
\min_i \left\lbrace v_{ij}, j \in J' \right\rbrace \\
\end{array}
\right.
\end{equation}
\begin{equation}
\label{eq:bigkey}
A^- = \left\lbrace v^-_1,\ldots,v^-_n \right\rbrace = \left\lbrace
\begin{array}{ll}
\min_i \left\lbrace v_{ij}, j \in J \right\rbrace \\
\max_i \left\lbrace v_{ij}, j \in J' \right\rbrace \\
\end{array}
\right.
\end{equation}
where \( J \) corresponds to benefit criteria and \( J' \) to cost criteria, and for \(  i=1,2,\ldots,m \) and \(  j=1,2,\ldots,n \).

    \item Step 5. Determine the separation measures of each alternative by calculating the separation distances from \( A^+ \) ( \( d^+_i \) ) and \( A^- \) ( \( d^-_i \) ) using the n-dimensional Euclidean distance method:
\begin{equation}
\label{eq:separationdistances}
d^+_i = \sqrt{\sum_{j=1}^{n}(v_{ij}-v^+_j)^2},
\end{equation}
\begin{equation}
\label{eq:separationdistances}
d^-_i = \sqrt{\sum_{j=1}^{n}(v_{ij}-v^-_j)^2}.
\end{equation}

    \item Step 6. Calculate the relative closeness to the ideal solution by determining the closeness coefficient for each alternative:
\begin{equation}
\label{eq:separationdistances}
R_i = \frac{d_i^-}{d_i^+ + d_i^-}.
\end{equation}

    \item Step 7. Rank the alternatives in descending order based on \( R_i \) values, where an alternative \( A_i \) closer to \( A^+ \) and further from \( A^- \) will have a higher ranking as \( R_i \) approaches 1.

\end{itemize}

Regarding the data, the 5 m/pixel Digital Elevation Models (DEMs) derived from LOLA data \cite{BARKER2021} were chosen for their high spatial resolution\footnote{\url{https://pgda.gsfc.nasa.gov/products/78}}, which is crucial for analyzing terrain features and visibility across the 13 candidate regions and the scientifically significant Site 004 \cite{NASA2020b}. The slope maps accompanying these DEMs provide essential information for assessing lander stability and traverse planning. The solar illumination, PSRs, and Earth sunlight reflection maps \cite{Mazarico2011Icar}, with a resolution of 240 m/pixel, were included to evaluate potential landing sites with regard to energy availability and volatile retention\footnote{\url{https://imbrium.mit.edu/BROWSE/EXTRAS/ILLUMINATION/}}. The 1 $\mu$m absorption depth band map, obtained from M$^3$ data at 1.4 km/pixel \citep{Moriarty2024JGRE}, was selected to provide insights into mineralogy of the explorable area. Lastly, the 1:5,000,000 scale geologic map, derived from SELENE Kaguya and LOLA data \cite{Fortezzo2020}, was used to ensure comprehensive geologic context, with spatial resolutions of 60 m/pixel and 100 m/pixel, respectively.

To handle these maps, we utilize two GIS tools: \textit{ArcGIS} \cite{ArcGIS2024} and \textit{Global Mapper} \cite{GMapper2024}. We utilize Bresenham's line algorithm to generate visibility maps considering the lunar curvature, which include one map representing the total area visible from the HLS windows and another for the line of sight of astronaut-HLS within the 2 km radius limited by the mission requirements \cite{Coan2020}.

The starting point involves analyzing LOLA topography images \cite{BARKER2021} of the 14 regions, where potential lunar landing locations will be evaluated and prioritized. This analysis begins by defining the parameters established for the Artemis III HLS \cite{NASA2019}, which include a landing precision of 100 m from any target landing site and slope requirements with vertical orientations ranging between 0 and 5$^{\circ}$. We generate a buffer zone around all pixels with a slope of less than 5$^{\circ}$ in the LOLA topography maps. This ensures that if a lunar module lands on any pixel within the delineated polygons, it will have a safety margin of at least 100 m up to a pixel with a slope greater than 5$^{\circ}$. We note that the reported geolocation uncertainty is approximately 10–20 cm horizontally and 2–4 cm vertically per pixel. However, that around 90\% of the 5 m/pixel data is interpolated, meaning it does not consist of directly measured values.

\section{Results}\label{decision problem}
\subsection{Generation of the potential landing locations (alternatives)}\label{alternatives} 

\begin{figure*}[h!]
\centering
\includegraphics[width=1\textwidth]{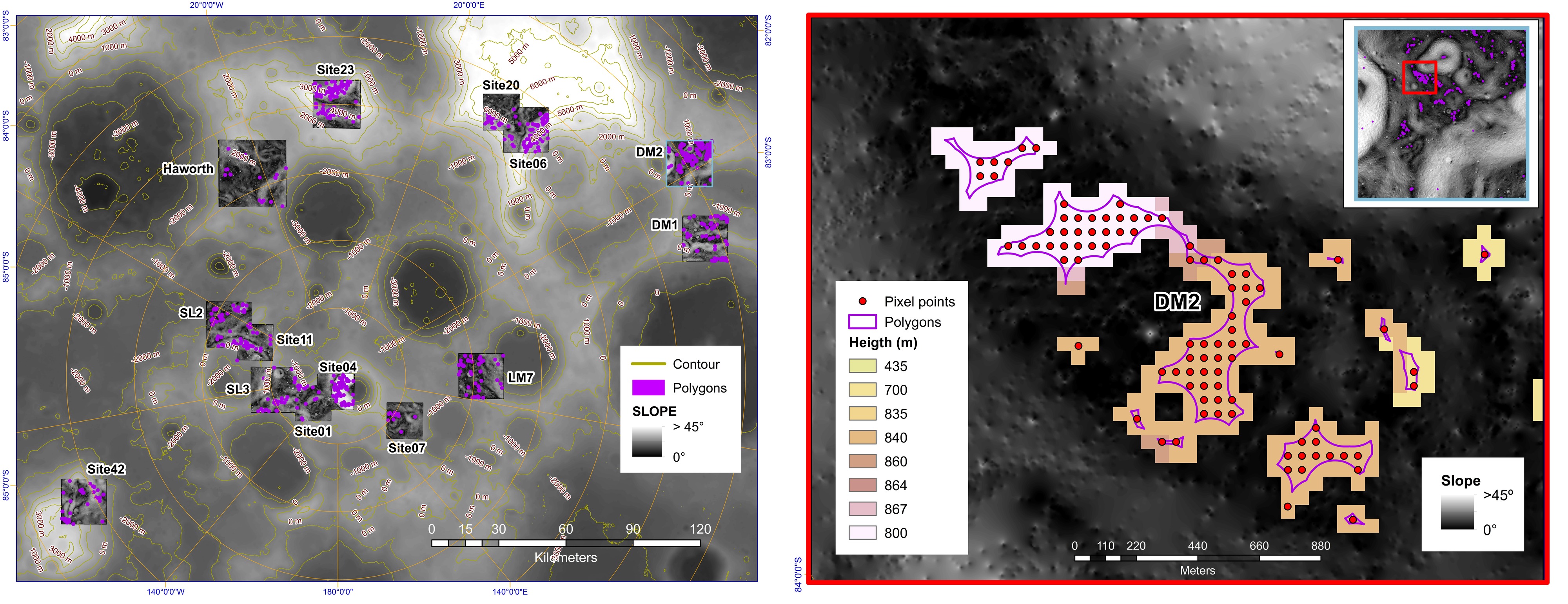}
\caption{Left: All polygons that meet HLS landing requirements for slope compliance. Contours represent 1 km elevation intervals. Concentric rings at steps of 1$^{\circ}$ in latitude. Right: Example of the polygons and pixel point (alternatives; landing locations) within the DM2 site.}\label{fig_polygons}
\end{figure*}

The buffer applied results in 732 polygons that meet the aforementioned slope conditions. The areas of these polygons range between 25 m$^2$ and 25,000 m$^2$. Figure \ref{fig_polygons} (left) shows the polygons that meet the buffer requirements along with concentric rings of 1$^{\circ}$ in latitude. Due to computational constraints, we select all pixel points within the polygons imposing a resolution of 50 m/pixel. Figure \ref{fig_polygons} (right) shows an example of a more detailed visualization of the Site DM2 is provided. Each of the points located inside each polygon corresponds to a potential landing location, constituting one of the alternatives to be evaluated. Considering all the points, there are a total of 1247 alternatives to rank in the decision problem. Table \ref{table1} lists the number of locations by region, together with the new official name of each region.

\begin{table}[h!]
\centering
\caption{Number of points per region. We include the new official name of each region.}
\begin{tabular}{lcc}
\hline
Region & Official name & Points \\
\hline
DM1           & Amundsen Rim &  149        \\ 
DM2           & Nobile Rim 2 &    473        \\ 
Haworth       & Haworth &  20        \\ 
LM7           & Faustini Rim A &  80        \\ 
Site01        & Connecting Ridge &  20        \\ 
Site04        & &  56        \\ 
Site06        & Nobile Rim 1 &  94        \\ 
Site07        & Peak Near Shackleton &  19        \\ 
Site11        & de Gerlache Rim 1 &   15        \\ 
Site20        & Leibnitz Beta Plateau &  23        \\ 
Site23        & Malapert Massif &  102        \\ 
Site42        & de Gerlache-Kocher Massif &  56        \\ 
SL2           & de Gerlache Rim 2 &  104       \\ 
SL3           & Connecting Ridge Extension &   36        \\ 
TOTAL         & &  1247         \\ 
\hline
\end{tabular}
\label{table1}
\end{table}

\subsection{Criteria to evaluate the potential landing locations}\label{criteria} 

There are few studies that specifically analyze the selection criteria for lunar exploration missions, and those that exist differ not only in their study area but also in the mission focus (e.g., crewed, robotic rovers). A brief review of the existing literature addressing these criteria follows.

In some cases, a single criterion dominates the decision-making process for selecting high-priority lunar landing sites. For instance, \cite{KringDurda2012} identified the Schrödinger basin as a prime location due to the wide range of science and exploration objectives that could be achieved in this area \cite{Allenderetal2019}.

For missions targeting the rim of the lunar South Pole-Aitken basin, multiple parameters are considered. These include overall illumination percentage, temporal ground station contact, the size of the area with favorable conditions, and proximity to key basin features \cite{Koebeletal2012}. Other studies in the same region emphasize scientific criteria, such as the diversity of rock types representing large areas, the presence of crystalline impact-melt rocks, and basalt flow deposits \cite{Jolliffetal2010}.

Additional parameters, such as the stability of volatiles, PSRs, slope, and temperature conditions, have been used to prioritize landing sites. For example, \cite{Lemelinetal2014} identified five regions near the lunar north pole and seven near the south pole with high scientific potential based on these criteria.

Common criteria for lunar polar rover missions include a combination of sunlight availability, traversable terrain, subsurface volatiles, and communication capabilities. The selection of Haworth Crater exemplifies this approach, as it offers a favorable intersection of these factors \cite{Heldmanetal2016}. Similarly, in the Sverdrup-Henson crater region, criteria such as terrain slope, water ice presence, energy sources, and communication capabilities with Earth were considered for site selection \cite{Leoneetal2023}. Recently, studies have identified landing areas at the lunar south pole with smooth topography, favorable illumination, moderate temperatures, and volatile material presence as key features for dynamic exploration missions involving rovers or flybys \cite{Fengetal2024}. In addition to NASA, other countries as China has also analyzed the selection of landing sites for their series of Chang' e missions. Scientific and engineering constraints such as potential for resources utilization, gravity, magnetic field and other environmental features, topographic slope, terrain obstacles, communication ability and temperature, among others, were the main criteria which had influence in the assessment of landing sites \cite{Liuetal2020}.

For the Artemis III mission, candidate regions must not only meet HLS requirements but also satisfy additional criteria such as Earth communication capabilities, engineering feasibility, and adequate illumination \cite{Huangetal2023}. These and other relevant criteria are incorporated into our multi-criteria analysis, each of which will be detailed in the following sections.

\subsubsection{Criterion C$_{1}$.- Total visibility}

\begin{figure}[h!]
\centering
\includegraphics[width=1\columnwidth]{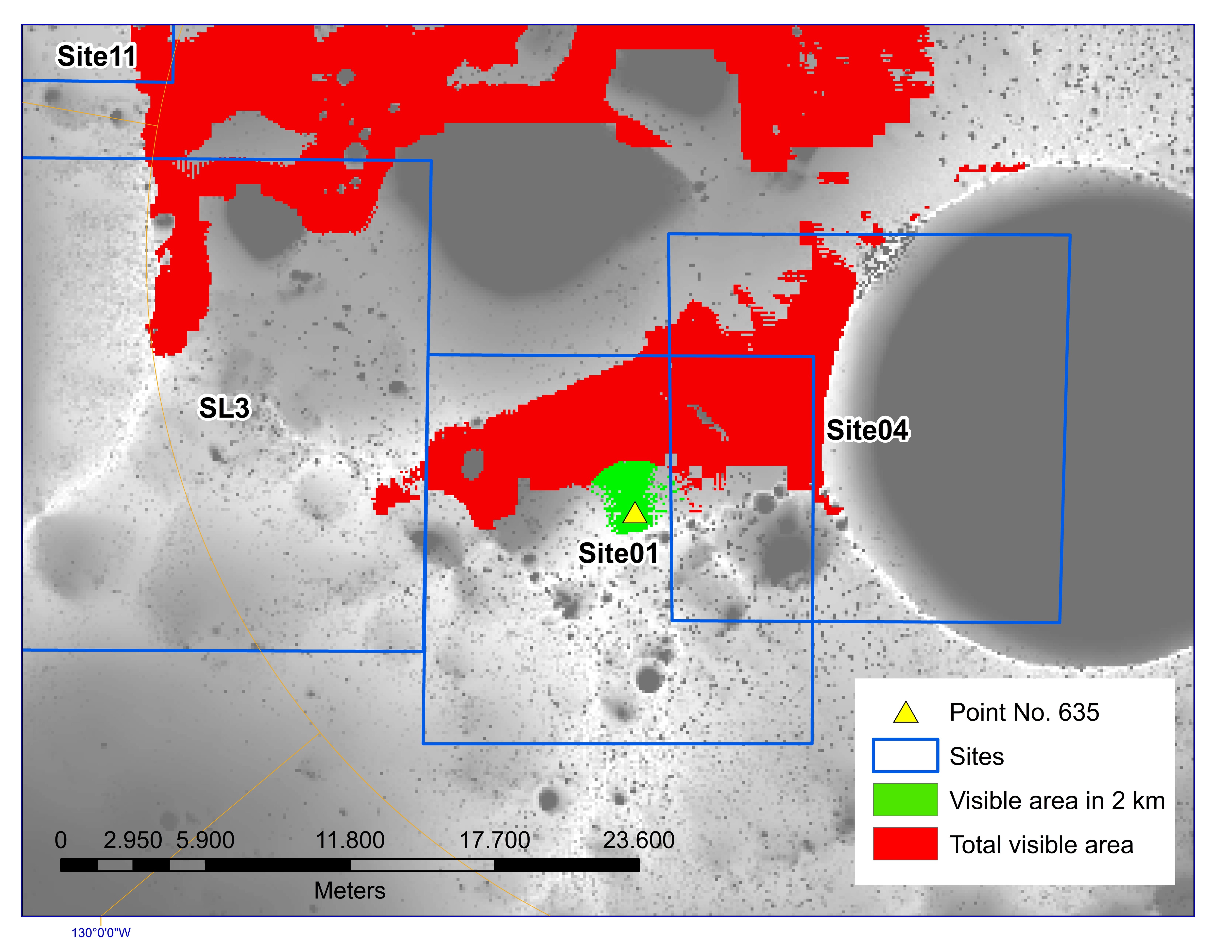}
\caption{Visible area from point number 635 located in Site 01. The area in red color represents the total visibility from that point. The area in green color shows the percentage of visibility for a 2 km radius.}\label{fig7}
\end{figure}

The first criterion is the total visibility of the lunar surface from the lander, which can provide scientific opportunities for intra-vehicular activities. Visibility from each point is evaluated using the method described in Section \ref{methodology} and the most updated lunar topography \cite{BARKER2021}. The objective is to identify the surface areas visible or hidden from an observer located at a height of 40 meters in the windows of the HLS \cite{Watson-Morgan2022}. Lunar curvature is considered, as the lunar horizon is about 12 km from flat terrain, but the furthest visible areas in our analysis are at distances of up to 200 km. The total visible area, measured in square kilometers, constitutes this criterion. Figure \ref{fig7} shows the result of the visible areas in red for a point located on Site SL2.

\subsubsection{Criterion C$_{2}$.- Explorable area visibility}

One feature that could facilitate astronauts' EVA and enhance their safety is maintaining a direct line of sight to the lander for reference. Therefore, we compute the percentage of the explorable walking surface, defined by a 2 km radius constrained by mission requirements, where the lander remains visible. The percentage of surface visible within this 2 km radius is the value of this criterion. Figure \ref{fig7} illustrates the explorable visible area in green for a point located on Site SL2.

\subsubsection{Criterion C$_{3}$.- PSRs and solar illumination}

\begin{figure}[h!]
\centering
\includegraphics[width=1\columnwidth]{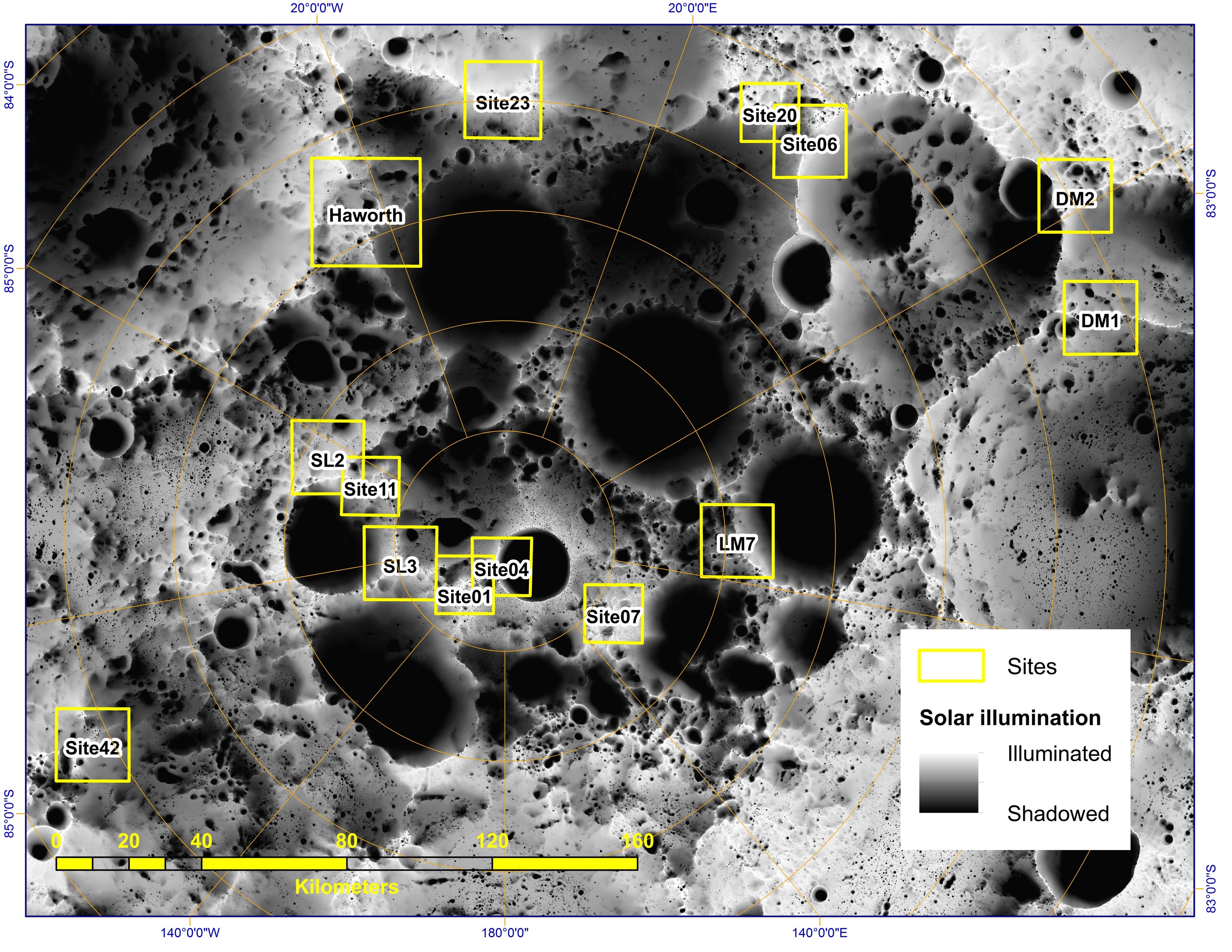}
\caption{Average solar illumination of the AEZ. Concentric rings at steps of 1$^{\circ}$ in latitude.}\label{fig9}
\end{figure}

This criterion evaluates both the incident solar illumination \cite{Mazarico2011Icar} at the potential landing site, critical for energy supply, and the possibility of exploring PSRs with astronauts during EVAs, important for the mission's success. Both features are derived from the illumination map. We seek a balance between these two parameters, as an area with perfect illumination but no PSRs to visit, or vice versa, is of limited interest. Therefore, we multiply these two parameters, ensuring that if one is deficient, it overrides the other. Thus, the value of this criterion is the product of the average illumination and the percentage of PSR within the 2 km radius explorable by astronauts. Figure \ref{fig9} shows the average solar illumination of the AEZ.

\subsubsection{Criterion C$_{4}$.- Direct communication with Earth}

\begin{figure}[h!]
\centering
\includegraphics[width=1\columnwidth]{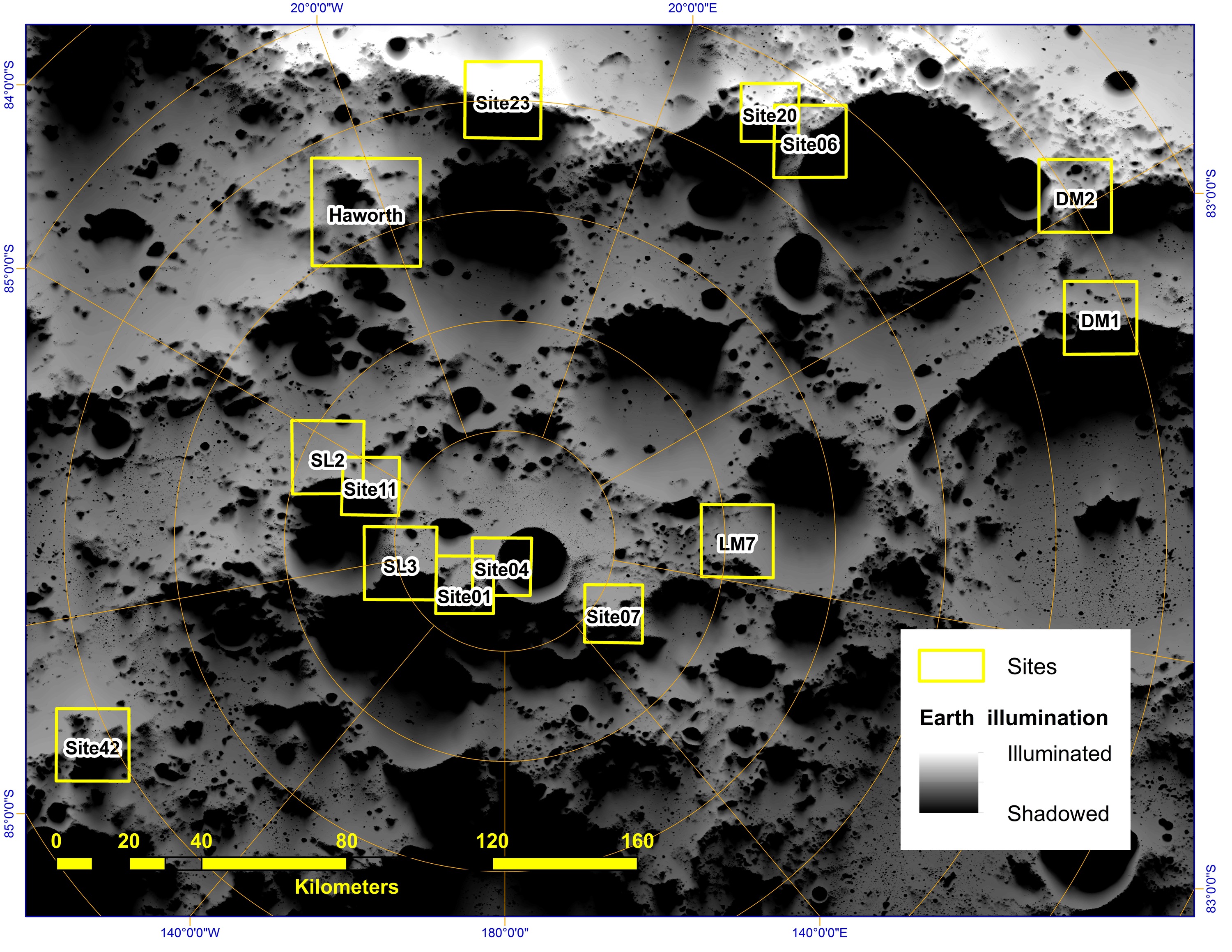}
\caption{Sunlight reflected by Earth on the AEZ indicates the direct communication of the lunar surface with Earth. Concentric rings at steps of 1$^{\circ}$ in latitude.}\label{fig10}
\end{figure}

This criterion assesses the potential for direct communication between the lunar surface and Earth by evaluating sunlight reflected by Earth \cite{Mazarico2011Icar} onto the AEZ. The level of reflected sunlight indicates the site's ability to maintain a line of sight with Earth. Figure \ref{fig10} shows the illumination map used.

\subsubsection{Criterion C$_{5}$.- Geological units in explorable area}

\begin{figure}[h!]
\centering
\includegraphics[width=1\columnwidth]{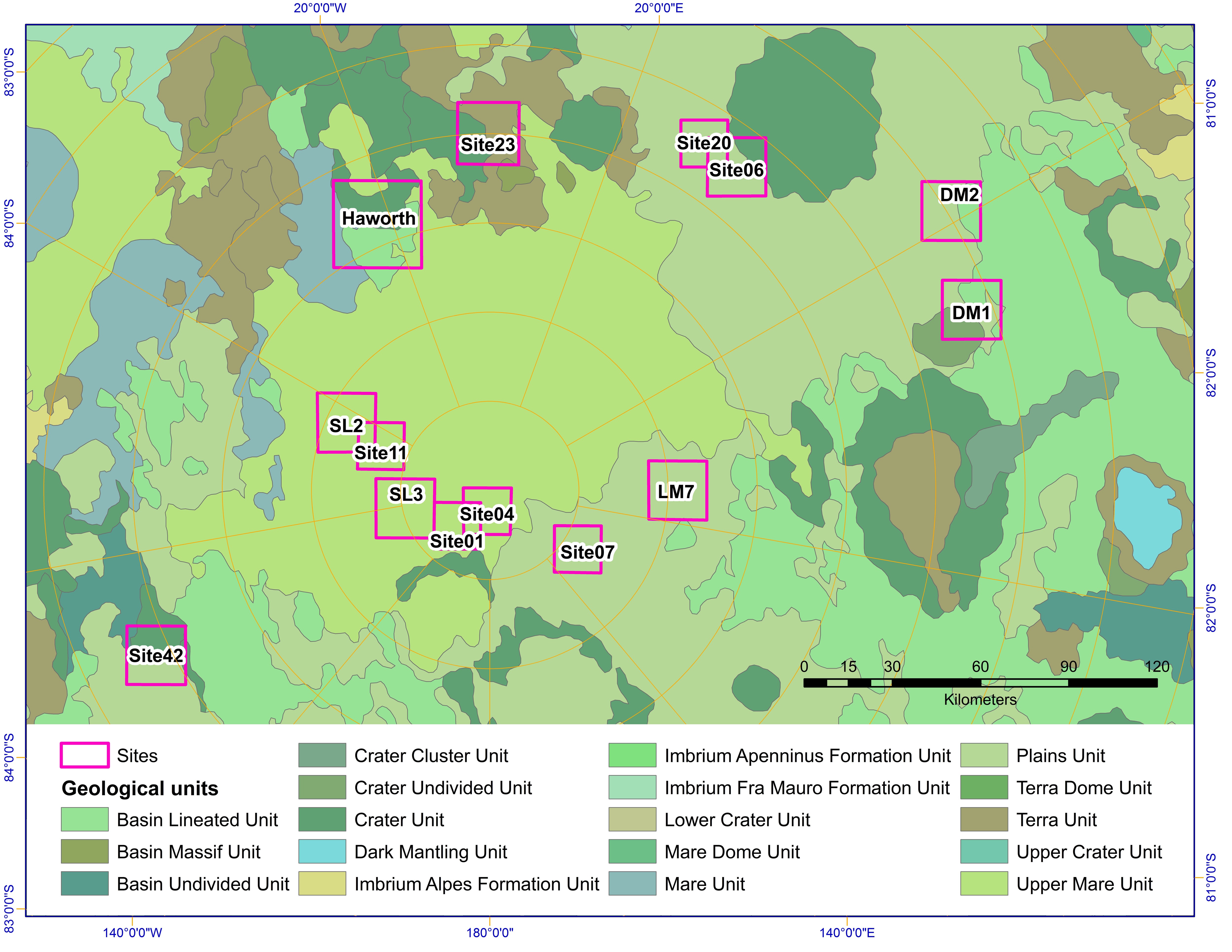}
\caption{Geological units of the AEZ. Concentric rings at steps of 1$^{\circ}$ in latitude.}\label{fig11}
\end{figure}

To enhance the geological return of the mission, this criterion evaluates the number of distinct geological units within a 2 km radius of the landing site. This assessment determines the variety of geological units that astronauts can explore during an EVA. The evaluation utilizes the Unified Geological Map of the Moon at a 1:5M scale \cite{Fortezzo2020}. Figure \ref{fig11} illustrates the map of geological units for the AEZ.

\subsubsection{Criterion C$_{6}$.- Mineralogy}

\begin{figure}[h!]
\centering
\includegraphics[width=1\columnwidth]{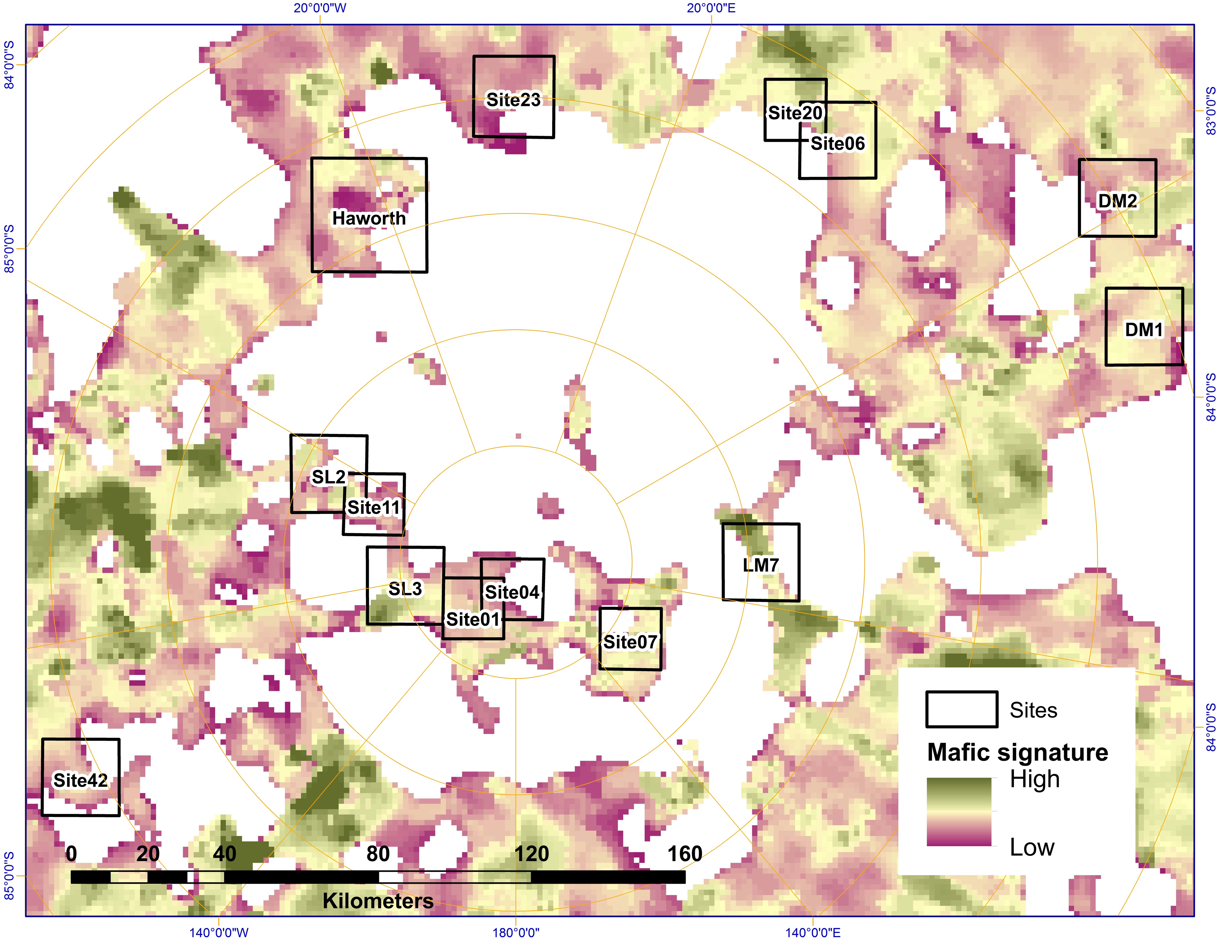}
\caption{Mafic mineral signature in the AEZ. 
Data from \cite{Moriarty2024JGRE}. Concentric rings at steps of 1$^{\circ}$ in latitude.}\label{fig12}
\end{figure}

While basalts rich in mafic minerals, such as those containing high proportions of magnesium and iron, are prevalent in lunar maria, the South Pole-Aitken basin is a significant source of mafic materials near the south pole, offering a distinct mafic signature in this region with high proportions of magnesium and iron. In the context of lunar exploration and \textit{in situ} resource utilization, these minerals are particularly valuable due to their composition, which includes elements useful for various applications. For example, mafic materials with higher olivine and pyroxene content melt at lower temperatures, simplifying processing, and they also contain more potentially extractable metallic phases. The most prominent mafic signatures are found in the South Pole-Aitken Basin, indicating the presence of impact melting and basin ejecta that can be sampled and characterized during the Artemis III mission \cite{Moriarty2024JGRE}. Utilizing data from the Moon Mineralogy Mapper on 1 $\mu$m band depths, it is possible to characterize mafic mineral abundances, as this absorption band is sensitive to the presence of pyroxene and olivine in a variety of lithologies, including those likely present in the South Pole-Aitken basin ejecta \cite{Moriarty2024JGRE}. Figure \ref{fig12} provides the mafic mineral signature map for the AEZ.

\subsection{Prioritization of the alternatives} \label{prioritization} 

Once the values for each criterion have been calculated for each potential landing location, the generated layers form a decision matrix comprising 1247 alternatives evaluated against 6 criteria. This matrix serves as the foundation for applying the TOPSIS multi-criteria analysis methodology, which facilitates the prioritization and selection of the most favorable areas for the lunar landing in the context of the Artemis III mission. The decision matrix is schematically represented in Table \ref{table2}. This decision matrix corresponds to stage 1 of the TOPSIS multi-criteria decision technique and therefore constitutes the starting point of the process of prioritization of alternatives according to methodology described in Section \ref{methodology}.

\begin{table}[h!]
\centering
\footnotesize
\caption{Decision matrix schema for 1247 alternatives evaluated against 6 criteria.}
\begin{tabular}{lccccccc}
\hline
ID   & Region    & C$_{1}$      & C$_{2}$    & C$_{3}$      & C$_{4}$       & C$_{5}$   & C$_{6}$   \\ \hline
1    & Site23  & 4240 & 79.86 & 0.00    & 24986 & 1 & 2.13 \\ 
2    & Site23  & 4112 & 80.02 & 0.00    & 0     & 2 & 0.84 \\ 
3    & Site20  & 75   & 75.72 & 2632    & 15780 & 2 & 7.10 \\ 
4    & Site06  & 4077 & 67.12 & 0.00    & 13112 & 1 & 5.58 \\ 
5    & Site06  & 4077 & 67.12 & 0.00    & 13112 & 1 & 5.58 \\ 
6    & Site06  & 4302 & 69.03 & 0.00    & 13471 & 1 & 6.35 \\ 
7    & Site06  & 3234 & 80.10 & 0.00    & 0     & 1 & 4.63 \\ 
8    & Site06  & 1189 & 89.18 & 0.00    & 0     & 1 & 4.42 \\ 
9    & DM2     & 393  & 54.78 & 4736    & 13053 & 1 & 4.62 \\ 
...  & ...     & ...     & ...   & ...     & ...      & ...  & ...  \\ 
1247 & Haworth & 1624 & 82.89 & 2668 & 13391 & 1 & 3.33 \\ \hline
\end{tabular}
\label{table2}
\end{table}

\begin{figure*}[h!]
\centering
\includegraphics[width=0.75\textwidth]{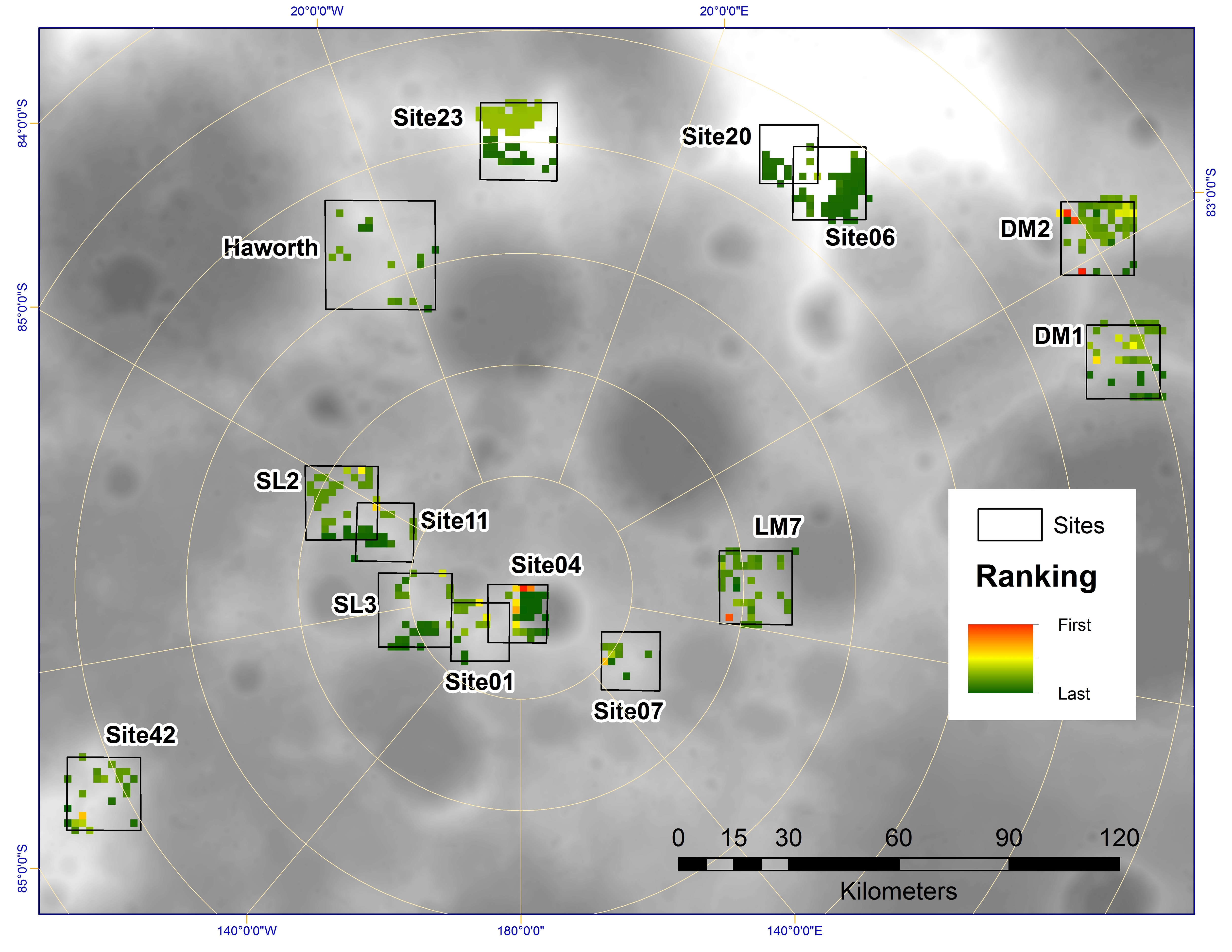}
\caption{Map of the potential lunar landing sites according to the Ranking TOPSIS. Concentric rings at steps of 1$^{\circ}$ in latitude.}\label{fig13}
\end{figure*}

By executing each stage of the TOPSIS algorithm with equal weighting for all criteria, we generate a ranking of alternatives based on their relative closeness to the ideal solution, R$_i$. Figure \ref{fig13} shows the TOPSIS ranking of potential lunar landing sites using a color-coded map. The Site DM2 (Nobile Rim 2) features several locations ranked at the top. Table \ref{table3} lists the 20 best locations in descending order according to the TOPSIS methodology, including their coordinates and values for each criterion. The optimal location, identified as point no. 358, is situated at the DM2 site. We note that Site DM2 exhibits an excellent combined engineering suitability, as analyzed by \cite{Huangetal2023}.

\begin{figure*}[h!]
\centering
\includegraphics[width=1\textwidth]{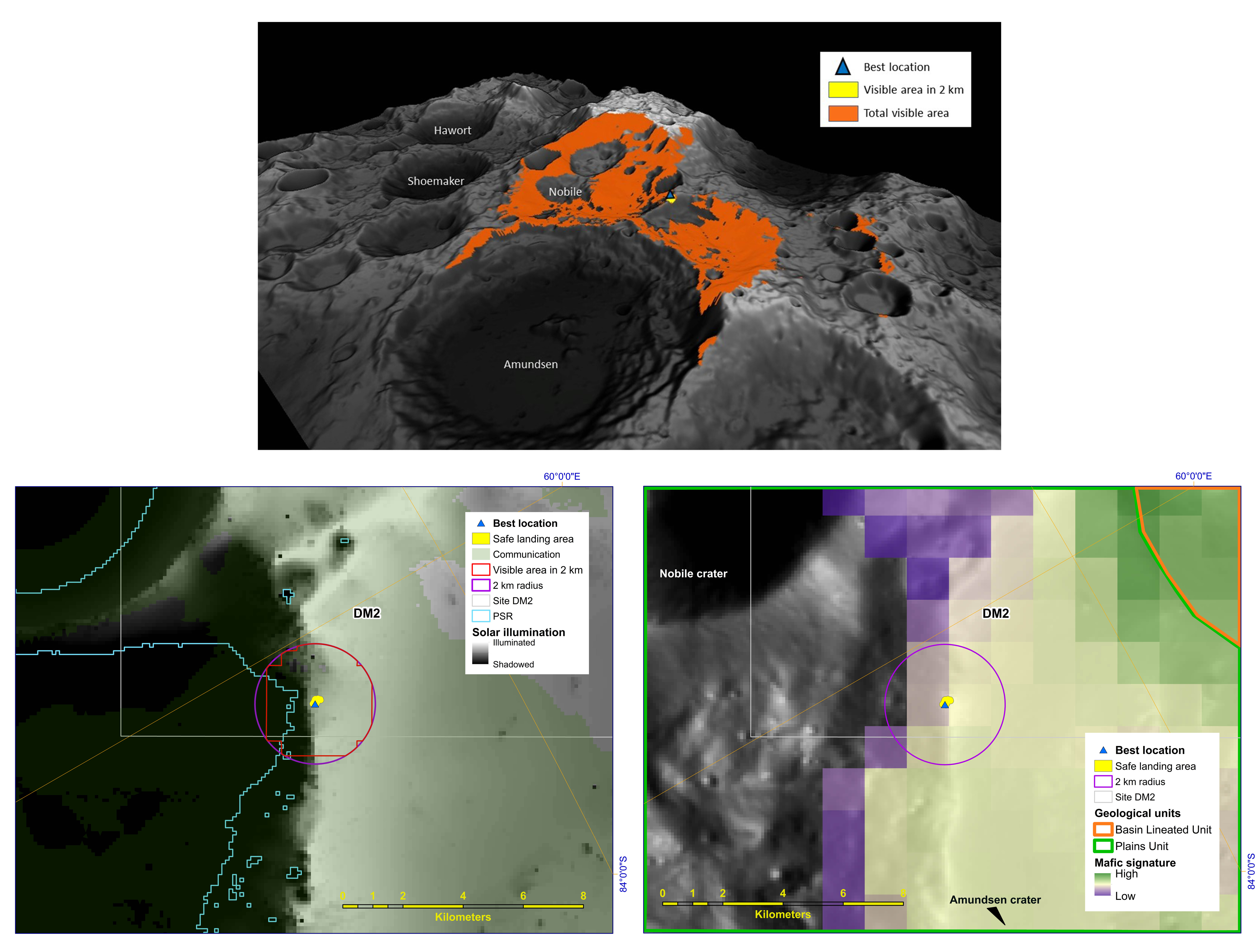}
\caption{Best location map (point no. 358 located in DM2 site). Top: 3D local topography (vertical exaggeration of x2.5) and visibility from the HLS window. Bottom-left: Direct communication with Earth, visible explorable area, PSRs, and solar illumination. Bottom-right: geological units and mafic mineral signature,}\label{fig_bestlocation}
\end{figure*}

Figure \ref{fig_bestlocation} provides detailed information regarding the criteria for this location. Analysis of the top 100 locations reveals that 36\% are in Site DM2, 30\% in Site 23, and 13\% in Site DM1, highlighting these as the most promising lunar landing sites. In contrast, no top 100 locations are found in Sites SL3, 20, and Haworth. The fact that the best location is within the DM2 candidate region could have been anticipated to some extent, as DM2 has the largest surface area meeting the HLS landing requirements among all candidate regions, and therefore the highest number of potential landing points.

Specifically, the optimal point for the lunar landing is located at latitude 84°12'5.61" S (-84.20156°) and longitude 60°41'59.61" E (60.69989°), within Site DM2. It exhibits medium mafic abundance and, like most locations, includes one explorable geological unit. It maintains direct communication with Earth for one-third of the time and offers the best ratio of explorable PSR and solar illumination at the landing point. Additionally, half of the explorable area retains direct sight with the HLS, and it is among the locations with the most visible total lunar surface.

\begin{table*}[h!]
\centering
\small
\caption{Ranking of the 20 best locations according to the TOPSIS methodology. All criteria have equal weight.}
\begin{tabular}{lcccccccccc}
\hline
ID & Region & R$_i$ (TOPSIS) & Latitude & Longitude & C$_{1}$ & C$_{2}$ & C$_{3}$ & C$_{4}$ & C$_{5}$ & C$_{6}$ \\ \hline
358         & DM2            & 0.6916               & 84°12'5.61"S                             & 60°41'59.61"E                             & 4067.5      & 44.8        & 140592.7    & 8315        & 1           & 3.70        \\ 
356         & DM2            & 0.6869               & 84°12'2.71"S                            & 60°41'8.51"E                              & 4067.5      & 44.8        & 138178.2    & 7746        & 1           & 3.70        \\ 
357         & DM2            & 0.6711               & 84°11'57.55"S                            & 60°41'37.19"E                             & 4067.5      & 44.8        & 129545.1    & 7746        & 1           & 3.70        \\ 
354         & DM2            & 0.6684               & 84°11'59.82"S                           & 60°40'17.42"E                             & 4026.2      & 44.5        & 128637.8    & 7746        & 1           & 3.70        \\ 
350         & DM2            & 0.6610               & 84°11'56.91"S                            & 60°39'26.35"E                             & 4026.2      & 44.5        & 126065.2    & 7016        & 1           & 3.70        \\ 
355         & DM2            & 0.6487               & 84°11'54.65"S                           & 60°40'46.11"E                             & 4026.2      & 44.5        & 120518.5    & 7746        & 1           & 3.70        \\ 
352         & DM2            & 0.6455               & 84°11'46.59"S                            & 60°40'23.74"E                             & 4115.9      & 41.3        & 119584.2    & 6935        & 1           & 3.70        \\ 
351         & DM2            & 0.6408               & 84°11'51.75"S                            & 60°39'55.05"E                             & 4026.2      & 44.5        & 118202.8    & 7016        & 1           & 3.70        \\ 
353         & DM2            & 0.6220               & 84°11'41.43"S                            & 60°40'52.41"E                             & 4115.9      & 41.3        & 111646.1    & 6935        & 1           & 3.70        \\ 
1121        & DM1            & 0.5653               & 84°11'30.63"S                           & 67°13'44.55"E                             & 313.7       & 49.6        & 108843.5    & 13506       & 1           & 4.59        \\ 
1122        & DM1            & 0.5649               & 84°11'30.78"S                            & 67°13'10.37"E                             & 313.7       & 49.6        & 110778.4    & 13506       & 1           & 3.88        \\ 
995         & DM2            & 0.5353               & 84°5'2.79"S                              & 55°33'15.48"E                             & 1968.4      & 31.5        & 104013.3    & 0           & 1          & 2.60        \\ 
661         & Site04         & 0.5294               & 89°57'33.08"S                          & 135°49'6.44"E                              & 327.1       & 60.2        & 115220.9    & 0           & 1           & 0.20        \\ 
998         & DM2            & 0.4916               & 84°2'16.50"S                             & 56°50'50.63"E                             & 994.2       & 43.2        & 88687.8     & 13587       & 1           & 3.22        \\ 
1223        & Site07         & 0.4689               & 89°0'18.22"S                           & 125°18'33.49"E                             & 2346.7      & 94.9        & 81886.9     & 12997       & 1           & 0.78        \\ 
744         & LM7            & 0.4567               & 88°5'13.01"S                            & 98°21'19.76"E                             & 362.4       & 82.6        & 88951.3     & 9121        & 1           & 0.00        \\ 
1043        & DM2            & 0.4373               & 83°46'3.73"S                            & 56°39'32.63"E                             & 449.1       & 69.8        & 73931.0     & 12981       & 1           & 4.52        \\ 
657         & Site04         & 0.4179               & 89°55'18.86"S                            & 78°44'15.06"E                             & 323.9       & 57.3        & 82422.2     & 0           & 1           & 0.17        \\ 
726         & SL2            & 0.3907               & 88°30'56.38"S                           & 61°10'2.35"O                              & 804.4       & 74.3        & 67061.3     & 14271       & 2           & 0.24        \\ 
1005        & Site07         & 0.3714               & 89°0'42.35"S                            & 131°16'44.46"E                             & 527.1       & 97.1        & 61877.1     & 10233       & 1           & 2.93        \\ \hline
\end{tabular}
\label{table3}
\end{table*}

\subsection{Sensitivity analysis}\label{sensitivity}

To analyze the robustness of our methodology, we modified the weights of the criteria, giving extra weight to those potentially more relevant for the Artemis III mission: high presence of PSRs and high solar illumination (C${3}$), and direct communication with Earth (C${4}$). Under these premises, we assigned a weight of 40\% to criterion C${3}$ and another 40\% to criterion C${4}$, distributing the remaining 20\% equally among the other criteria. We obtain a new ranking of alternatives by re-executing the TOPSIS algorithm with these new weights. Comparing the 20 best locations, we found that the top-ranked location (point 358 at the DM2 site) remained unchanged, and the top 19 locations were consistent, with only slight changes in their order starting from the ninth position (see Table \ref{table4}). We also note that C${5}$ has minimal impact on the ranking, as most locations allow access to only one geological unit. This sensitivity analysis confirms the robustness of the leading alternatives and validates the multi-criteria methodology applied.

\begin{table*}[h!]
\small
\centering
\caption{Ranking of the 20 best locations according to the TOPSIS methodology. Columns 3 and 4 consider all criteria with equal weight, while in columns 5 and 6, 80\% of the weight is assigned to PSRs and solar illumination (C${3}$) and direct communication with Earth (C${4}$).}
\begin{tabular}{lccccc}
\hline
 &                       & \multicolumn{2}{c}{All criteria with same weight} & \multicolumn{2}{c}{C$_{3}$ and C$_{4}$ with 80\% of weight} \\
ID    & Region & R$_i$ (TOPSIS)  & Ranking & R$_i$ (TOPSIS) & Ranking \\ \hline
358   & DM2    & 0.6916  & 1     & 0.8453  & 1     \\ 
356   & DM2    & 0.6869  & 2     & 0.8383  & 2     \\ 
357   & DM2    & 0.6711  & 3     & 0.8186  & 3     \\ 
354   & DM2    & 0.6684  & 4     & 0.8157  & 4     \\ 
350   & DM2    & 0.6610  & 5     & 0.8019  & 5     \\ 
355   & DM2    & 0.6487  & 6     & 0.7841  & 6     \\ 
352   & DM2    & 0.6455  & 7     & 0.7752  & 7     \\ 
351   & DM2    & 0.6408  & 8     & 0.7693  & 8     \\ 
353   & DM2    & 0.6220  & 9     & 0.7366  & 11    \\ 
1121  & DM1    & 0.5653  & 10    & 0.7476  & 10    \\ 
1122  & DM1    & 0.5649  & 11    & 0.7587  & 9     \\ 
995   & DM2    & 0.5353  & 12    & 0.6629  & 13    \\ 
661   & Site04 & 0.5294  & 13    & 0.7131  & 12    \\ 
998   & DM2    & 0.4916  & 14    & 0.6212  & 14    \\ 
1223  & Site07 & 0.4689  & 15    & 0.5754  & 16    \\ 
744   & LM7    & 0.4567  & 16    & 0.6084  & 15    \\ 
1043  & DM2    & 0.4373  & 17    & 0.5232  & 18    \\ 
657   & Site04 & 0.4179  & 18    & 0.5411  & 17    \\ 
726   & SL2    & 0.3907  & 19    & 0.4807  & 19    \\ 
1005  & Site07 & 0.3714  & 20    & 0.4365  & 22    \\ \hline
\end{tabular}
\label{table4}
\end{table*}

Additionally, we compare our results against three MCDC methods that have distinct theoretical foundations and computational procedures: MABAC \cite{PAMUCAR2015}. VIKOR \cite{Opri1998}, and CoCoSo \cite{Yazdani2019}. MABAC constructs a normalized decision matrix, each attribute is then adjusted by subtracting the border approximation area, and the alternatives are ranked based on these adjusted values. VIKOR considers the proximity to the ideal solution and the maximum group utility, balancing between the majority's and the minority's preferences. CoCoSo determines weighted normalized decision matrices, computing the sum of weighted normalized performance ratings, and calculating the compromise score. Table \ref{table_MCDCcomparison} shows the comparison of the ranking position for the best 25 potential landing locations, as evaluated by each method.

The results reveal a high degree of consistency in the top rankings across the four methodologies. Specifically, the best and second-best alternatives are consistently prioritized, indicating a strong consensus on the top-performing options and enhancing confidence in their suitability. However, as we move down the rankings, the similarity among the methods begins to diminish. While the top positions show remarkable agreement, the lower-ranking alternatives exhibit more variability. Notably, MABAC, VIKOR, and CoCoSo agree on the last position. Conversely, the alternative ranked last by TOPSIS is classified as fifth or sixth by the other three methods.

\begin{table}[h!]
\small
\centering
\begin{tabular}{lcccccc}
\hline
ID   & Region         & TOPSIS     & MABAC      & CoCoSo     & VIKOR      \\ \hline
358  & DM2          & 1          & 1          & 1          & 1          \\ 
356  & DM2          & 2          & 2          & 2          & 2          \\ 
357  & DM2          & 3          & 3          & 3          & 4          \\ 
354  & DM2          & 4          & 4          & 4          & 5          \\ 
350  & DM2          & 7          & 5          & 6          & 6          \\ 
355  & DM2          & 5          & 6          & 7          & 7          \\ 
352  & DM2          & 9          & 8          & 9          & 9          \\ 
351  & DM2          & 8          & 7          & 8          & 8          \\ 
353  & DM2          & 10         & 10         & 10         & 10         \\ 
1121 & DM1          & 14         & 12         & 12         & 12         \\ 
1122 & DM1          & 17         & 17         & 14         & 17         \\ 
995  & DM2          & 22         & 22         & 22         & 22         \\ 
661  & Site04       & 23         & 23         & 23         & 23         \\ 
998  & DM2          & 18         & 19         & 18         & 19         \\ 
1223 & Site07       & 15         & 11         & 11         & 11         \\ 
744  & LM7          & 21         & 21         & 21         & 21         \\ 
1043 & DM2          & 13         & 16         & 13         & 16         \\ 
657  & Site04       & 25         & 25         & 25         & 25         \\ 
726  & SL2          & 19         & 9          & 5          & 3          \\ 
1005 & Site07       & 20         & 18         & 16         & 18         \\ 
398  & DM1          & 11         & 14         & 19         & 14         \\ 
637  & Site04       & 24         & 24         & 24         & 24         \\ 
658  & Site04       & 16         & 20         & 20         & 20         \\ 
399  & DM1          & 12         & 15         & 15         & 15         \\ 
876  & DM1          & 6          & 13         & 17         & 13         \\ \hline
\end{tabular}
\caption{Comparative ranking position of the best 25 potential landing locations using different MCDC methods.}
\label{table_MCDCcomparison}
\end{table}

\section{Conclusions} \label{conc}

This study reaffirms the effectiveness of integrating GIS software with MCDM methodologies for addressing prioritization problems. The application of the TOPSIS technique reveals that the optimal candidate region for the Artemis III landing is located in Site DM2 (Nobile Rim 2). Specifically, the optimal point is at latitude 84°12'5.61" S (-84.20156°) and longitude 60°41'59.61" E (60.69989°). This potential landing location meets the HLS requirements and demonstrates the best balance among all possible points across all regions regarding visibility, PSRs, solar illumination, direct communication with Earth, geological units, and abundance of mafic materials. Sensitivity analysis shows that this location is robust, remaining the best despite readjusting the criteria weights to prioritize lighting, PSRs, and communication with Earth, as well as using other MCDC methods. In general terms, Site DM2 holds the best locations, followed by Site DM1. On the other hand, Sites SL3, 20, and Haworth do not possess any locations within the top 100. 

The flexibility of the approach allows for the addition or removal of criteria, adjustment of weightings, and application to new zones. However, computation time increases exponentially with higher resolution, particularly for visibility calculations, while the rest of the criteria and TOPSIS processing can be considered almost instantaneous.

As a result of the demonstrated applicability of the GIS-MCDM combination for lunar exploration, we propose to replicate this combination of tools and techniques on other celestial bodies with available cartographic information from remote sensors. Mars, given its prominence in upcoming space agency science goals, could be an excellent candidate. This approach would enable similar preliminary analyses for future Solar System exploration. 


\section*{Acknowledgement}

EP-A and JMS-L have carried out this work in the framework of the project Fundación Seneca (22069/PI/22), Spain. EP-A has received support from the LUMIO project funded by the Agenzia Spaziale Italiana (2024-6-HH.0). JMS-L thanks funding from the Ministerio de Ciencia e Innovación of Spain (grants PID2020-112754GB-I00 and PID2021-128062NBI00). 

\bibliography{mybibfile}

\end{document}